\documentclass[12pt]{article}
\usepackage{amsfonts}
\begin{document}
\centerline{\large\bf BITWISTOR FORMULATION OF SPINNING PARTICLE
\footnote{Contribution to the Proceedings of the International
Workshop ``Supersymmetry and Quantum Symmetries'' (SQS'03, Dubna,
24-29 July 2003)}} \vspace{0.2cm} \centerline{\large {\bf S.
Fedoruk}$^\dag$ {\large {\bf and}} {\bf V.G. Zima}$^*$}
\centerline{$^\dag$ {\small\it Ukrainian Engineering--Pedagogical
Academy, Kharkov, Ukraine}} \centerline{$^*$ {\small\it Kharkov
National University, Kharkov, Ukraine}}
\begin{abstract}
Twistorial formulation of a particle of arbitrary spin has been
constructed. Equations of twistor transformation are obtained for
massive and massless spinning particles. Twistor space of the
massive particle is formed by two twistors and two complex
scalars. In the massive case, integral transformations relating
twistor fields with usual space--time fields have been
constructed.
\end{abstract}
\section{Introduction}
Penrose twistors~\cite{PenMac}, \cite{Pen} are a powerful tool for
the analysis of (super)symmetric models of point--like and
extended objects. In the twistor approach mainly massless
(super)par\-ticles~\cite{Sorokin} have been considered so far. In
twistor formalism massive particle, especially with non--zero
spin, is investigated in a rather limited number of
paper~\cite{PenMac}, \cite{Perj}-\cite{FedZim2}. But even in case
of massless spinning particle the connection of standard twistor
formulation and formulation in (real) space-time produces series
of questions.

In the twistor formulation phase space of a massless particle is
described by two canonically conjugated each other Weyl spinors
$\lambda_\alpha$, $\bar\lambda_{\dot\alpha}=
(\overline{\lambda_\alpha})$ and $\bar\omega^{\alpha}$,
$\omega^{\dot\alpha} =(\overline{\bar\omega^{\alpha}})$ which
combined in four--component Penrose twistor $Z_a
=\left(\lambda_\alpha\, ,\,\,\, \omega^{\dot\alpha}\right)$. In
twistor variables Lagrangian of massless particle has the form
\begin{equation}\label{lagr-tw0}
L=-{\textstyle \frac{1}{2}}(\dot{\bar Z}^a Z_a -\bar Z^a \dot Z_a)
-N ({\textstyle \frac{i}{2}}{\bar Z}^a Z_a +j)
\end{equation}
where conjugate twistor is defined as $\bar Z^{a}=
\left(\bar\omega^{\alpha}\, ,\,\,\,
-\bar\lambda_{\dot\alpha}\right)$ and $N$ is Lagrange multiplier
for twistor spin constraint
\begin{equation}\label{spin-s0}
{\textstyle \frac{i}{2}}{\bar Z}^a Z_a +j={\textstyle
\frac{i}{2}}(\bar\omega^{\alpha}\lambda_\alpha -
\bar\lambda_{\dot\alpha}\omega^{\dot\alpha})+j\approx 0\,.
\end{equation}

Obtained by Noether procedure conserved charges
$P_{\alpha\dot\alpha}=\lambda_\alpha \bar\lambda_{\dot\alpha}$,
$M_{\alpha\beta}=i\lambda_{(\alpha}\bar\omega_{\beta )}$, $\bar
M_{\dot\alpha\dot\beta}=i\bar\lambda_{(\dot\alpha}\omega_{\dot\beta
)}$, $K^{\dot\alpha\alpha}=\omega^{\dot\alpha}
\bar\omega^{\alpha}$, $D={\textstyle
\frac{1}{2}}(\bar\omega^{\alpha}\lambda_\alpha +
\bar\lambda_{\dot\alpha}\omega^{\dot\alpha})$ corresponding to
conformal transformations give for pseudovector Pauli-Lubanski
$W_{\alpha\dot\alpha}=P_{\alpha}{}^{\dot\beta}\bar
M_{\dot\beta\dot\alpha}-P^{\beta}{}_{\dot\alpha} M_{\beta\alpha}$
following value $ W_{\alpha\dot\alpha}=(-{\textstyle
\frac{i}{2}}{\bar Z}^a Z_a)P_{\alpha\dot\alpha}$. Thus the
classical consideration shows that model with twistor
Lagrangian~(\ref{lagr-tw0}) describes massless particles of finite
spin ($P^2=0$, $W^2=0$). Spin (helicity) is defined by quantity
$(-{\textstyle \frac{i}{2}}{\bar Z}^a Z_a)$, which equal on
classical level constant $j$ due to constraint~(\ref{spin-s0}).
Therefore constant $j$ is `classical helicity' of massless
particle.

Fundamental relations of twistor transformation
\begin{equation}\label{inc-s0}
p_{\alpha\dot\alpha}=\lambda_\alpha
\bar\lambda_{\dot\alpha}\,;\qquad \quad \omega^{\dot\alpha}=
{\textstyle \frac{1}{2}}x^{\dot\alpha\alpha}\lambda_\alpha\,
,\qquad \bar\omega^{\alpha}={\textstyle
\frac{1}{2}}\bar\lambda_{\dot\alpha} x^{\dot\alpha\alpha}
\end{equation}
connect twistor variables and space--time ones
$p_{\alpha\dot{\alpha}}=p_\mu \sigma^\mu_{\alpha\dot{\alpha}}$,
$x^{\dot{\alpha}\alpha}=x^\mu \sigma_\mu^{\dot{\alpha}\alpha}$.
But in traditional writing of the relations~(\ref{inc-s0}) it
should be isotropy condition for twistor $ {\bar Z}^a Z_a
=\bar\omega^{\alpha}\lambda_\alpha -
\bar\lambda_{\dot\alpha}\omega^{\dot\alpha}=0\,. $ Since `twistor
$Z_a$ of general form describes classical massless particle with
helicity $-{\textstyle \frac{i}{2}}{\bar Z}^a Z_a$' (\cite{Pen},
Sec.6, \S\,2, text after (6.3.9)) the isotropic twistor describes
massless particle of zero helicity $j=0$. Thus basic
conditions~(\ref{inc-s0}) determine the correspondence between
twistor formulation~(\ref{lagr-tw0}) and space--time one of
massless particle only at zero spin $j=0$. Massless particles with
nonzero helicity are obtained in quantum spectrum upon the
transition from space--time formulation to twistor formulation
either by introducing nonzero spin (helicity) constant $j$
directly into the spin constraint $\bar Z Z -2ij\approx 0$ in the
twistor action~(\ref{lagr-tw0}) `by hand' or taking into account
an ordering ambiguity of quantum variables in the spin constraint.
Taking into account basic ideology of twistor approach as
alternative for space--time description~\cite{Pen}, from set of
formulations of spinning particle it must be space--time
formulation corresponding to one~(\ref{lagr-tw0}) at fulfilment of
basic relation for twistor transformation which generalize the
expressions~(\ref{inc-s0}).

Twistor description of massive particle requires with necessity
more than one twis\-tor~\cite{PenMac}, \cite{Perj}-\cite{FedZim2}.
It follows directly from this that time--like momentum of massive
particle can be resolved only with using two or more twistor
spinors $p_{\alpha\dot\alpha}=
\lambda_{\alpha}^i\bar\lambda_{\dot\alpha i}$. In a minimal case
two twistors ($i=1,2$) are used. But the complete twistor
description of the massive particle of an arbitrary spin has not
been accomplished yet. It should include the constraints and the
Lagrangian of the massive spinning particle and also a correct
chooses of corresponding variables in twistor formulation. In
contrast to the massless case where, in principle, the twistor
description of a particle of arbitrary helicity is achieved by
using only one twistor in the massive case it is necessary to use
some spinning variables in addition to the twistor ones
\footnote{Otherwise the constraints and the Lagrangian of the
system would be nonlinear and rather complicated [private
discussion with J.Lukierski].}. In this case we have some analog
with case of massless superparticle where for description of
nonzero superhelicity it is necessary to use additional variables
which together with twistor variables form supertwistor. As in
case massless particle it is necessary to have corresponding
space--time formulation of massive spinning particle which
connected with twistor formulation after using of fundamental
relations of twistor transformation. It is important also that
massless case and massive one have certain analogies. In ideal
massless case can be obtained from massive one in level zero mass
$m\rightarrow 0$.

In present paper we follow the constructive way to finding of
twistor formulation of spinning particle implies using the
appropriate space--time formulation. For this aim from all
space--time formulations the more appropriate formulation are
those in which the spin degrees of freedom are described by means
of commuting variables. Also from such formulations there are
appropriate ones in which spin variables are spinors (for
obtaining arbitrary spins including half--integer ones) and
describing of arbitrary spins is realized in uniform way. The
formulation relativistic spinning particle with index
spinor~\cite{ZimFed2} is more appropriate formulation for these
aims. In this formulation uniform Lagrangian describes massive and
massless spinning particles. Also spinning particle with index
spinor has a some analogy with usual superparticle. Therefore for
our aim we can exploit the some elements of transition from
space--time formulation of superparticle to twistor one.
\section{Twistor formulation of massive spinning particle}
We take index spinor formulation~\cite{ZimFed2} of spinning
particle as starting point for obtaining twistor formulation. In
index spinor formalism spinning particle is described with
space--time vector $x^\mu$ and commuting Weyl spinor
$\zeta^\alpha$. In first order formalism its Lagrangian has the
form~\cite{ZimFed2}
\begin{equation}\label{lagr-s}
L= -{\textstyle \frac{1}{2}}
p_{\alpha\dot\alpha}\dot\Pi^{\dot\alpha\alpha} + {\textstyle
\frac{1}{2}} V (p_{\alpha \dot\alpha}p^{\dot\alpha\alpha}-2m^2) -
\Lambda (\zeta^\alpha p_{\alpha \dot\alpha}\bar\zeta^{\dot\alpha}
-j)\, ,
\end{equation}
where the bosonic `superform' is
$\Pi^{\dot\alpha\alpha}\equiv\dot\Pi^{\dot\alpha\alpha}d\tau
\equiv dx^{\dot\alpha\alpha}
+2i\bar\zeta^{\dot\alpha}d\zeta^\alpha -
2id\bar\zeta^{\dot\alpha}\zeta^\alpha$, $p_\mu$ is momentum vector
of particle with mass $m$ and real scalars $V$ and $\Lambda$ are
Lagrange multipliers. After quantization the wave function of the
spinning particle $\Psi(\zeta,\bar\zeta)=e^{-\zeta\hat
p\bar\zeta}\Phi(\zeta)$ is expressed by holomorphic polynomial
$\Phi(\zeta)=\zeta^{\alpha_1}\ldots\zeta^{\alpha_s}\phi_{\alpha_1\ldots\,\alpha_s}$
on spinor $\zeta$. The spin (helicity) of particle $s$ is equal to
the constant $j$ renormalized by ordering constants\footnote{When
potential term $-\Lambda(\zeta\hat p\bar\zeta -j)$ in
Lagrangian~(\ref{lagr-s}) is absent after quantization the wave
function is expressed by function which is polynomial series on
$\zeta$, $\bar\zeta$ with non-fixed degree of homogeneity
$$
\Phi(\zeta,\bar\zeta)=\sum_{n=0}^\infty \sum_{m=0}^\infty
\zeta^{\alpha_1}\ldots\zeta^{\alpha_n}\bar\zeta^{\dot\alpha_1}\ldots\bar\zeta^{\dot\alpha_m}
\phi_{\alpha_1\ldots\,\alpha_n\dot\alpha_1\ldots\,\dot\alpha_m}\,.
$$
In this case we have in spectrum the set of particles with all
arbitrary spins (helicities). Thus index spinor variables have the
role analogical to role of the spinor variables in unfolded
formulation of higher spin field theory~\cite{Vas}.}.

Twistor formulation of massive spinning particle, which
corresponds to the space--time formulation~(\ref{lagr-s}), has
been constructed in~\cite{FedZim3}. For this we introduce pure
gauge variables in initial system which are Lorentz
harmonics~\cite{GIKOS}, \cite{Band}, \cite{ZimFed1}. In what
follows after canonical transformation we exclude space--time
variables by means of gauge fixing and remain only with twistor
variables. Canonical transformations and gauge fixing conditions
have physical meaning. In particular they produce the fundamental
conditions of twistor transformations relating the space--time
formulation and twistor one.

The massive spinning particle in the twistor formulation is
described by the variables $\lambda_\alpha^i$,
$\bar\lambda_{\dot\alpha i} = \overline{(\lambda_\alpha^i)}$,
$\omega^{\dot\alpha i}$, $\bar\omega^{\alpha}_i
=\overline{(\omega^{\dot\alpha i})}$; $\xi^i$, $\bar\xi_i
=\overline{(\xi^i)}$; $i=1,2$. The connection of twistor variables
with the space--time ones is defined by the conditions of twistor
transformation
\begin{equation}\label{p-tw-tr}
p_{\alpha\dot\alpha}=\lambda_\alpha^i \bar\lambda_{\dot\alpha i}\,
;
\end{equation}
\begin{equation}\label{o-tw-tr}
\bar\omega^{\alpha}_i ={\textstyle
\frac{1}{2}}\bar\lambda_{\dot\alpha i} x^{\dot\alpha\alpha} +
i\bar\xi_i\zeta^\alpha \, , \qquad \omega^{\dot\alpha i}
={\textstyle \frac{1}{2}}x^{\dot\alpha\alpha}\lambda_{\alpha}^i -
i\xi^i \bar\zeta^{\dot\alpha}\, ;
\end{equation}
\begin{equation}\label{x-tw-tr}
\xi^i =\zeta^\alpha \lambda_\alpha^i \, ,\qquad \bar\xi_i =\bar
\lambda_{\dot\alpha i}\bar\zeta^{\dot\alpha} \, .
\end{equation}
The twistor phase space of massive spinning particle is subjected
to the set of the first class constraints
\begin{equation}\label{c-h}
M\equiv \lambda^{\alpha i}\lambda_{\alpha i}
+\bar\lambda_{\dot\alpha i} \bar\lambda^{\dot\alpha i} + 4m\approx
0 \, ,
\end{equation}
\begin{equation}\label{c-D}
D_i{}^j \equiv {\textstyle \frac{i}{2}}\left(
\bar\omega^{\alpha}_i \lambda_\alpha^j - \bar\lambda_{\dot\alpha
i} \omega^{\dot\alpha j} \right) + \bar\xi_i \xi^j \approx 0 \, ,
\quad S \equiv \bar\xi_i \xi^i - j\approx 0 \, .
\end{equation}

Spinors $\lambda^i$ and $\omega^i$ are components of the twistors
$Z_a^i =\left(\lambda_\alpha^i\, ,\,\,\, \omega^{\dot\alpha
i}\right)$. Introducing in a standard way conjugate twistors $\bar
Z_{\dot a i} =\overline{(Z_a^i)} = \left(\bar\lambda_{\dot\alpha
i}\, ,\,\,\, \bar\omega^{\alpha}_i \right)$, $\bar Z^{a}_i =
g^{a\dot b} \bar Z_{\dot b\,i} = \left(\bar\omega^{\alpha}_i\,
,\,\,\, -\bar\lambda_{\dot\alpha i} \right)$, the kinetic terms
are rewritten as $-{\textstyle
\frac{1}{2}}p_{\alpha\dot\alpha}\Pi^{\dot\alpha\alpha} =
{\textstyle \frac{1}{2}}(\bar Z^a_i\, dZ_a^i -d\bar Z^a_i\, Z_a^i)
+ i(d\bar\xi_i\, \xi^i -\bar\xi_i\, d\xi^i )$. The twistor
mass--shell constraints~(\ref{c-h}) take the form $ M= Z_a^i
I^{ab} Z_{bi} +\bar Z^a_i I_{ab} \bar Z^{bi} +4m \approx 0$ where
$ I^{ab}$, $I_{ab}$ are asymptotic twistors~\cite{Pen}. Also the
constraints~(\ref{c-D}) are represented as covariant contractions
of twistors $D_i{}^j = {\textstyle \frac{i}{2}} \bar Z^a_i Z_a^j +
\bar\xi_i \xi^j \approx 0$. Thus the twistor formulation of
massive spinning particle is described by Lagrangian
\begin{equation}\label{lagr-betwist}
L = {\textstyle \frac{1}{2}}\left(\bar Z^a_i\, \dot{Z}_a^i
-\dot{\bar Z}^a_i\, Z_a^i \right) -i\left(\dot{\bar\xi}_i\, \xi^i
-\bar\xi_i\, \dot{\xi}^i \right) - \Lambda_m M-\Lambda_j{}^i
D_i{}^j- \Lambda_s S\, ,
\end{equation}
where $\Lambda_m$, $\Lambda_j{}^i$, and $\Lambda_s$ are Lagrange
multipliers for constraints~(\ref{c-h}), (\ref{c-D}).
\section{Twistorial formulation of massless particle}
In level of zero mass $m\rightarrow 0$ we can leave only one
spinor $\lambda_\alpha$ which resolves the light--like vector of
four--momentum $p_\mu$. The equation of twistor
transformation~(\ref{p-tw-tr})-(\ref{x-tw-tr}) for massless
particle take the form
\begin{equation}\label{inc-s}
p_{\alpha\dot\alpha}=\lambda_\alpha \bar\lambda_{\dot\alpha}\, ;
\quad \bar\omega^{\alpha} ={\textstyle
\frac{1}{2}}\bar\lambda_{\dot\alpha} x^{\dot\alpha\alpha} +
i\bar\xi\zeta^\alpha\, ; \qquad \xi =\zeta^\alpha \lambda_\alpha
\end{equation}
and c. c. Here $\lambda_\alpha$, $\omega^{\dot\alpha}$ and c. c.
are spinor components of twistors $Z_a =(\lambda_\alpha\, ,\,
\omega^{\dot\alpha})$ and $\bar Z^{a} = (\bar\omega^{\alpha}\, ,\,
-\bar\lambda_{\dot\alpha})$ as whereas commuting complex scalar
$\xi$, $\bar\xi=\overline{(\xi)}$ is corresponding spin variable
in twistor formalism. The phase space of the massless particle is
subject to the first class constraints
\begin{equation}\label{spin-s}
D_0 \equiv i( \bar\omega^{\alpha} \lambda_\alpha -
\bar\lambda_{\dot\alpha} \omega^{\dot\alpha} ) + 2\bar\xi \xi =
i\bar Z^a Z_a + 2\bar\xi \xi \approx 0 \, ,\qquad S \equiv \bar\xi
\xi - j\approx 0 \, .
\end{equation}
These constraints follow from the constraints of massive
twistorial particle~\cite{FedZim3} in case of using one twistor
and zero mass.

Thus in twistor variables the massless particles with nonzero
helicity is described by Lagrangian
\begin{equation}\label{lagr-tw-s}
L ={\textstyle \frac{1}{2}}(\bar Z^a\, \dot{Z}_a -\dot{\bar Z}^a\,
Z_a ) -i(\dot{\bar\xi}\, \xi -\bar\xi\, \dot{\xi} ) -\Lambda D_0-
\Lambda_s S\, .
\end{equation}
This system is the space--time formulation~(\ref{lagr-s}) of the
spinning particle reformulated in the twistor approach. Thus after
using conditions~(\ref{inc-s}) kinetic terms of the
Lagrangian~(\ref{lagr-s}) transform in kinetic terms of the
Lagrangian~(\ref{lagr-tw-s}) $-{\textstyle
\frac{1}{2}}p_{\alpha\dot\alpha}\Pi^{\dot\alpha\alpha} =
{\textstyle \frac{1}{2}}(\bar Z^a\, dZ_a -d\bar Z^a\, Z_a) +
i(d\bar\xi\, \xi -\bar\xi\, d\xi )$. Resolving the condition
$p_{\alpha\dot\alpha}= \lambda_\alpha \bar\lambda_{\dot\alpha}$
and using the equations~(\ref{inc-s}) for scalars $\xi$,
$\bar\xi$, the spin constraint $\zeta\hat p\bar\zeta -j\approx 0$
of the Lagrangian~(\ref{lagr-s}) takes the form of the constraint
$S\approx 0$ in twistor variables.

We can exclude the variables $\xi$ and $\bar\xi$ with the help of
first class constraint $\bar\xi \xi -j\approx 0$. After that
massless particle is described by Lagrangian~(\ref{lagr-tw0}). Due
to the spin constraint ${\textstyle \frac{i}{2}}\bar Z^a Z_a +j
\approx 0$ in Lagrangian~(\ref{lagr-tw0}) the twistor $Z_a$ is
nonisotropic (even on classical level) and describes massless
particle with nonzero helicity. We obtain the twistor formulation
with the nonisotropic twistor upon transition from the space--time
formulation because of the presence of the second terms in the
incidence conditions~(\ref{inc-s}). The presence in the incidence
conditions, written in real (not complex!) space--time, of the
second terms for spinning case is well known~\cite{Pen}. Real rays
which correspond to the twistor $Z=(\lambda ,\omega)$ with the
incidence conditions~(\ref{inc-s}) (nonisotropic twistor) form the
Robinson congruence.

Note that the conditions~(\ref{inc-s}) for definition of spinors
$\omega$ can be represented as twistorial shift
$\omega^{\dot\alpha} \rightarrow \omega^{\dot\alpha} -
i\xi\bar\zeta^{\dot\alpha}$, $\bar\omega^{\alpha} \rightarrow
\bar\omega^{\alpha} + i\bar\xi\zeta^{\alpha}$ along spinor
$\zeta$. Since $\zeta$ and $\lambda$ are orthogonal,
$\zeta\lambda\neq 0$, we obtain variation of helicity ${\textstyle
\frac{i}{2}}
(\bar\lambda_{\dot\alpha}\omega^{\dot\alpha}-\bar\omega^\alpha
\lambda_\alpha) \rightarrow {\textstyle
\frac{i}{2}}(\bar\lambda_{\dot\alpha}\omega^{\dot\alpha}-\bar\omega^\alpha
\lambda_\alpha) +j$. This twistorial shift is distinguished from
twistor  shift~\cite{SSTV} $\omega^{\dot\alpha} \rightarrow
\omega^{\dot\alpha} +l \bar\lambda^{\dot\alpha}$,
$\bar\omega^{\alpha} \rightarrow
\bar\omega^{\alpha}+l\lambda^{\alpha}$ along spinor $\lambda$
where $l$ is a length constant. That shift results in a
modification of particle (or string) interactions with background
fields~\cite{SSTV} and does not produce any change of helicity
since under this shift
$\bar\lambda_{\dot\alpha}\omega^{\dot\alpha}-\bar\omega^\alpha
\lambda_\alpha ={\rm inv}$.

The variables of the twistorial formulation~(\ref{lagr-tw-s}),
coordinates of twistor $Z_a =(\lambda_\alpha
,\omega^{\dot\alpha})$ and complex scalar $\xi$, may be combined
in quantity ${\cal Z}_{\cal A} =(Z_a ;\xi ) = (\lambda_\alpha
,\omega^{\dot\alpha}; \xi )$ which has five of complex components
and can be called as `bosonic supertwistor'. Introducing conjugate
quantities $\bar{\cal Z}_{\dot{\cal A}} =(\bar Z_{\dot a} ;\bar\xi
) = (\bar\lambda_{\dot\alpha}, \bar\omega^{\alpha} ; \bar\xi )$,
$\bar{\cal Z}^{{\cal A}} = g^{{\cal A}\dot{\cal B}} \bar{\cal
Z}_{\dot{\cal B}}= (\bar Z^{a} ;-2i\bar\xi ) =
(\bar\omega^{\alpha}, -\bar\lambda_{\dot\alpha} ; -2i\bar\xi )$ we
see that Lagrangian~(\ref{lagr-tw-s}) without last term
$-\Lambda_s (\bar\xi \xi -j)$ takes the form ${\textstyle
\frac{1}{2}} ( \bar{\cal Z}^{{\cal A}}\dot{\cal Z}_{{\cal A}} -
\dot{\bar{\cal Z}}^{{\cal A}} {\cal Z}_{{\cal A}}) - \Lambda
\bar{\cal Z}^{{\cal A}}{\cal Z}_{{\cal A}}$.

In the variables $\alpha$, $\beta$, $\gamma$, $\delta$ introduced
by $\lambda_1 =i(\alpha +\beta )$, $\lambda_2 =i(\gamma +\delta
)$, $\omega^1 = \alpha -\beta$, $\omega^2 = \gamma -\delta $ the
norm of `bosonic supertwistor' ${\textstyle \frac{i}{2}}\bar{\cal
Z}^{{\cal A}}{\cal Z}_{{\cal A}} = {\textstyle \frac{i}{2}}
(\bar\omega^\alpha \lambda_\alpha -
\bar\lambda_{\dot\alpha}\omega^{\dot\alpha}) +\bar\xi \xi$ takes
the form $ {\textstyle \frac{i}{2}}\bar{\cal Z}^{{\cal A}}{\cal
Z}_{{\cal A}} = \bar\beta\beta +\bar\delta\delta +\bar\xi\xi
-\bar\alpha\alpha -\bar\gamma\gamma $ and is quadratic Hermitian
form of signature $(+++--)$. Thus the Lagrangian~(\ref{lagr-tw-s})
without last term describing the set of particles with all
possible helicities is invariant under global transformations of
group $U(3,2)$ which play `bosonic superconformal' group in
twistor formalism. At transition to space--time
formulation~(\ref{lagr-s}) these $U(3,2)$--transformations reduce
to nonlinear transformation including usual the conformal
transformations, the `bosonic supersymmetric' transformations and
`bosonic superboosts'.
\section{Wave function of the twistorial spinning particle}
Canonical quantization a la Dirac of massive spinning particle in
the twistor formulation has been carry out in~\cite{FedZim3}. The
wave function is $(2J+1)$--component field $\Psi_M (\lambda
,\bar\lambda )$, defined up to local transformations acting on
index $M=-J,-J+1,...,J$
\begin{equation} \label{su2-trans}
\Psi_M^\prime (\lambda^\prime) ={\bf D}^J_{MN}(h) \Psi_N
(\lambda)\, ,
\end{equation}
where $h \in SU(2)$ and $\lambda^{\prime i}
_\alpha=h^i_j\lambda^j_\alpha$. The ${\bf D}^J_{MN}$ is matrix of
$SU(2)$--transformations of weight $J$. Thus the wave function is
defined in fact on homogeneous space ${\cal M} =G/H
=SL(2,C)/SU(2)$. In form of $SU(2)$-index $i=1,2$ the index $M$ is
collective index $M=(i_{1}\ldots i_{2J})$. Then the wave function
which represented twistor field of massive spinning particle is
\begin{equation} \label{wave-tw}
\Psi_M(\lambda,\bar\lambda) =\Psi_{i_{1}\ldots\,
i_{2J}}(\lambda,\bar\lambda)\, ,
\end{equation}
The wave function is completely symmetric $\Psi_{i_{1}\ldots\,
i_{2J}}=\Psi_{(i_{1}\ldots\, i_{2J})}$.

The relation of the usual space--time spin--tensor fields
$\Phi_{\alpha_{1}\ldots\alpha_{2J}}(x)$ with the twistor
fields~(\ref{wave-tw}) is established by means of an integral
transformation in the following way. One constructs
$SU(2)$-invariant expressions contracting the twistor fields
$\Psi_{i_{1}\ldots i_{2J}}(\lambda,\bar\lambda)$ with twistor
spinors
$\lambda_{\alpha_{1}}^{i_{1}},\ldots,\lambda_{\alpha_{2J}}^{i_{2J}}$.
Obtained expressions being Lorentz spin--tensors are defined on a
homogeneous space $SL(2,C)/SU(2)$. After integration with an
invariant measure $d^3\lambda$ of space $SL(2,C)/SU(2)$ with the
standard Fourier exponent $e^{ix^\mu p_\mu}$ where
$p_\mu=-{\textstyle
\frac{1}{2}}p_{\alpha\dot\alpha}\sigma^{\dot\alpha\alpha}_\mu=-{\textstyle
\frac{1}{2}}\lambda^i\sigma_\mu\bar\lambda_i$ we obtain usual
space--time fields\footnote{Analogous integral transformations for
massless twistor fields have been considered in~\cite{ZimFed1},
\cite{BandLuk}, \cite{PlSorTs}}
\begin{equation} \label{field}
\Phi_{\alpha_{1}\ldots\,\alpha_{2J}}(x)=\int d^3\lambda
e^{-{\textstyle \frac{i}{2}}x^\mu\lambda^k\sigma_\mu\bar\lambda_k}
\lambda_{\alpha_{1}}^{i_{1}}\ldots\lambda_{\alpha_{2J}}^{i_{2J}}
\Psi_{i_{1}\ldots\, i_{2J}}(\lambda,\bar\lambda)\, .
\end{equation}
These fields are totally symmetric in spinor indices
$\Phi_{\alpha_{1}\ldots\alpha_{2J}}(x)=\Phi_{(\alpha_{1}\ldots\alpha_{2J})}(x)$
and give us standard $(2J+1)$-component field description of
massive spin $J$. Due to the presence of the exponent in the
integral representation for the fields~(\ref{field}) they satisfy
automatically massive Klein--Gordon equation
$(\partial^\mu\partial_\mu-m^2)\,
\Phi_{\alpha_{1}\ldots\,\alpha_{2J}}(x)=0$.

We can obtain different, but equivalent descriptions of free
massive particles of spin $J$. The $SU(2)$-invariants can be
constructed by contraction the twistor fields $\Psi_{i_{1}\ldots
i_{2J}}(\lambda,\bar\lambda)$ only with twistor spinors
$\bar\lambda_{\dot\alpha}^{i}$. In this way we obtain the
space--time fields only with primed spinor indices. When
constructing the $SU(2)$-invariants, the part of the
$SU(2)$-indices in the twistor field $\Psi_{i_{1}\ldots
i_{2J}}(\lambda,\bar\lambda)$ can be contracted with spinors
$\lambda_{\alpha}^{i}$ whereas the remaining ones are contracted
with $\bar\lambda_{\dot\alpha}^{i}$. As result we obtain the
spin--tensor field
$\Phi_{\alpha_{1}\ldots\,\alpha_{n}}{}^{\dot\alpha_{n+1}\ldots\,\dot\alpha_{2J}}(x)$
with $n$ unprimed spinor indices and $2J-n$ primed ones. By
construction this field is symmetric with respect all unprimed and
all primed indices and satisfy massive Klein--Gordon equation and
transversal condition
$\partial^{\mu}\sigma_{\mu}{}_{\dot\alpha_{n+1}}^{\alpha_1}
\Phi_{\alpha_{1}\ldots\,\alpha_{n}}{}^{\dot\alpha_{n+1}\ldots\,\dot\alpha_{2J}}(x)=0$.
These fields at fixed $J$, describing the massive spin $J$
particle, are related to each other by Fierz--Pauli equations
$i\,\partial^\mu \,\sigma_\mu^{\dot\beta\alpha_{n}}\,
\Phi_{\alpha_{1}\ldots\,\alpha_{n}}{}^{\dot\alpha_{1}\ldots\,\dot\alpha_{2J-n}}=
m\,\Phi_{\alpha_{1}\ldots\,\alpha_{n-1}}{}^{\dot\alpha_{1}\ldots\,\dot\alpha_{2J-n}\dot\beta}$,
$i\,\partial^\mu \,\sigma_{\mu\beta\dot\alpha_{n}}\,
\Phi_{\alpha_{1}\ldots\,\alpha_{2J-n}}{}^{\dot\alpha_{1}\ldots\,\dot\alpha_{n}}=
m\,\Phi_{\alpha_{1}\ldots\,\alpha_{2J-n}\beta}{}^{\dot\alpha_{1}\ldots\,\dot\alpha_{n-1}}$.
United the fields
$\Phi_{\alpha_{1}\ldots\,\alpha_{n}}{}^{\dot\alpha_{n+1}\ldots\,\dot\alpha_{2J}}(x)$
with fixed $J$ in a unique fields we obtain description of spin
$J$ massive particle in Bargman--Wigner or Rarita-Schwinger
formalism.
\par\bigskip
{\bf Acknowledgements.} This work was supported in part by INTAS
Grant INTAS-2000-254 and by Ukrainian National Found of
Fundamental Researches under the Project \textsl{N} 02.07/383. We
would like to thank I.A.~Bandos, E.A.~Ivanov, S.O.~Krivonos,
J.~Lukierski, A.J.~Nurmagambetov, D.P.~Sorokin and M.A.~Vasiliev
for interest to this work and for many useful discussions.

\end{document}